\documentclass[aps,prl,reprint]{revtex4-1}
\usepackage[utf8x]{inputenc}
\usepackage{amsmath}
\usepackage{graphicx}

\usepackage{appendix}
\newcommand{\br}[1]{\left(#1\right)}
\newcommand{\sq}[1]{\left[#1\right]}
\begin{document}

\pagenumbering{gobble} 

{\bf Comment on ``Gradient Dynamics Description for Films of Mixtures and Suspensions: Dewetting Triggered by Coupled Film Height and Concentration Fluctuations''}

\vspace{10pt}

In the Letter \cite{Thiele2013} published by Thiele \emph{et al.}, 
the authors make use of a model first proposed by Clarke \cite{Clarke2005} 
for the description of a binary fluid film described by 
a free energy functional depending on material volume fraction $\phi(x)$ and film height $h(x)$. 
The model can be summarised (ignoring several terms common to both \cite{Thiele2013} and \cite{Clarke2005} that are not relevant to this discussion) by 
\begin{align} \label{eq:free_energy}
 F\sq{\phi(x),h(x)} = \int f(\phi,h) + hg(\phi) dx
\end{align}
where (using the terminology of \cite{Thiele2013}) $f(\phi, h)$ is the wetting energy and $g(\phi)$ is the bulk free energy. 
In order to describe the evolution of the film, both \cite{Thiele2013} and \cite{Clarke2005} used a gradient dynamics approach 
to derive a set of coupled equations for the evolution of the volume fraction and height of the film. 
Although Thiele \emph{et al.}'s model \cite{Thiele2013} reproduced the conclusions published by Clarke 
(namely that a binary mixture film will be less stable due to coupling of fluctuations of height and composition than if these fluctuations were not coupled \cite{Clarke2004} 
and that instabilities leading to dewetting can be triggered through this coupling \cite{Clarke2005, Thomas2010}) 
there is an important difference in the way the constraints of the system (conservation of height and material) 
are incorporated into the dynamics. 
%
%

The constraints for height and material can be written as $A^{-1}\int h(x) dx = h_0$ and $A^{-1}\int h(x) \phi(x) dx = h_0 \phi_0$ respectively, 
where $h_0$ and $\phi_0$ are the height and composition of the initially homogeneous film respectively, and $A$ is the area of the film. 
A gradient dynamics approach involves the gradient of functional derivatives of the free energy \eqref{eq:free_energy} 
with respect to the order parameters of the system (the underlying assumption is that the flux of material is proportional to the chemical potential). 
In \cite{Clarke2005} variations of Eq. \eqref{eq:free_energy} were performed with respect to the conserved order parameter $h(x)$ and the non-conserved order parameter $\phi(x)$, 
but Clarke used \emph{constrained} functional derivatives, as set out by G\'{a}l \cite{Gal2002, Gal2007}, to incorporate the constraints. 
The constrained functional derivatives used in Clarke's gradient dynamics were given by \cite{Clarke2005} 
\begin{align} 
%
%
  \mu_{K\phi} = \frac{\delta F}{\delta\phi} - \frac{h(x)}{A\phi_0h_0} \int \phi(x') \frac{\delta F}{\delta\phi} dx' \label{eq:Clarke_constraint_funcs1} \\
  \mu_{Kh} = \frac{\delta F}{\delta h} - \frac{\phi(x)}{A\phi_0h_0} \int \phi(x') \frac{\delta F}{\delta \phi} dx' \label{eq:Clarke_constraint_funcs2}
\end{align}
In \cite{Thiele2013} the functional derivatives variations of Eq. \eqref{eq:free_energy} were performed with respect to the conserved order parameters $h(x)$ and $h(x)\phi(x)$, 
yielding $\delta F / \delta h$ and $\delta F / \delta (h\phi)$. 

Note that the incorporation of constraints into functional derivatives is not to satisfy the constraints themselves, 
since the constraints are already achieved locally by the dynamics, but to ensure that the dynamics are correct. 
Checking linear stability limits is an ideal way in which one can test if the constraints have been incorporated properly. 
The simplest comparison of the dynamical equations obtained by Thiele \emph{et al.} and Clarke can be done by 
linearising their equations such that $h=h_0 + \delta h$, $\phi=\phi_0 + \delta \phi$. 
The integrals in Eqs. \eqref{eq:Clarke_constraint_funcs1} - \eqref{eq:Clarke_constraint_funcs2} become 
$h_0^{-1}\partial_\phi f \vert_{h_0,\phi_0}  + \partial_{\phi}g \vert_{h_0,\phi_0} $ 
and the Fourier transforms of the linearised equations of both Clarke and Thiele can both be written at lowest order in the wavevector $q$ as 
(see note in Ref. \cite{factor} for the constant $\mathcal{C}$ in Eq. \eqref{eq:linear2})
\begin{align}
  \br{\frac{3\eta}{h_0^3}} \frac{\partial \delta h_q}{\partial t} &= q^2 \sq{ \br{ \frac{\partial^2 f}{\partial h^2} }\delta h_q + L_{h \phi} \delta \phi_q } \label{eq:linear1}                  \\
  \mathcal{C} \frac{\partial \delta \phi_q}{\partial t} &= q^2 \sq{ L_{\phi h} \delta h_q + \br{ \frac{\partial^2 f}{\partial \phi^2} + h_0 \frac{\partial^2 g}{\partial \phi^2} }\delta \phi_q } \label{eq:linear2} \\ 
  L_{h \phi} &= L_{\phi h} = \br{ \frac{\partial^2 f}{\partial h \partial \phi} - \frac{1}{h_0}\frac{\partial f}{\partial \phi} }. \label{eq:linear3}
\end{align}

Contrary to the footnote Ref. 28 of \cite{Thiele2013}, 
we see that \cite{Thiele2013} and \cite{Clarke2005} both obtain the same limit in the linear regime 
and reproduce the required thermodynamic stability criterion, 
which is plotted as a phase diagram in Fig. 2 of \cite{Thiele2013} 
(this phase diagram can be obtained simply by applying the main result of \cite{Clarke2004} to the free energy \eqref{eq:free_energy}, 
without deriving and linearising any dynamical equations).

%
%
It appears that their are no qualitative differences in the conclusions of \cite{Thiele2013} and \cite{Clarke2005}. 
The simplicity of Thiele \emph{et al.}'s treatment 
(intuitively, one can even see how the stability criterion of \cite{Clarke2005} and \cite{Clarke2004} implies the form of the gradient dynamics given in \cite{Thiele2013}), 
and the fact that variations with respect to the two independent conserved quantities remains entirely local, 
suggests that performing variations with respect to conserved variables is sufficient to ensure constraints are honoured in gradient dynamics approaches to 
deriving dynamical equations. 


\vspace{10pt}

\noindent
\begin{minipage}{.5\textwidth}
Sam Coveney, Nigel Clarke

\indent
\quad {\small University of Sheffield,}

\quad {\small Hicks Building, Hounsfield Road, S3 7RH, UK}
\end{minipage}

\bibliographystyle{unsrt}


\end{document}